\documentclass[prb,preprint, groupaddress , amsmath, amssymb, showkeys, nofootinbib]{revtex4-1}
\usepackage{graphicx}
\usepackage{dcolumn}
\usepackage{bm}
\usepackage{upgreek}
\usepackage{times}
\usepackage{mathrsfs}
\usepackage{multirow}
\usepackage{fancyhdr} 
\usepackage{hyperref} 
\usepackage[version=4]{mhchem} 

\RequirePackage[normalem]{ulem}
\RequirePackage{color}\definecolor{RED}{rgb}{1,0,0}\definecolor{BLUE}{rgb}{0,0,1}

\begin{document}

\title{Bimolecular theory of non-radiative recombination in semiconductors with disorder}

\author{Oleg Rubel}
\email[]{rubelo@mcmaster.ca}
\affiliation{Department of Materials Science and Engineering, McMaster University, 1280 Main Street West, Hamilton, Ontario L8S 4L8, Canada}

\date{\today}

\begin{abstract}
The original Shockley-Read-Hall recombination statistics is extended to include recombination of localized excitations. The recombination is treated as a bimolecular process rather than a monomolecular recombination of excitons. The emphasis is placed on an interplay between two distinct channels of radiative recombination (shallow localized states \textit{vs} extended states) mediated by trapping of photogenerated charge carriers by non-radiative centers. Results of a numerical solution for a given set of parameters are complemented by an approximate analytical expression for the thermal quenching of the photoluminescence intensity in non-degenerate semiconductors derived in the limit of low pump intensities. The merit of a popular double-exponential empirical function for fitting the thermal quenching of the photoluminescence intensity is critically examined.
\end{abstract}

\maketitle

%
%
\section{Introduction}\label{Sec:Introduction}

\begin{figure}[h]
 \includegraphics{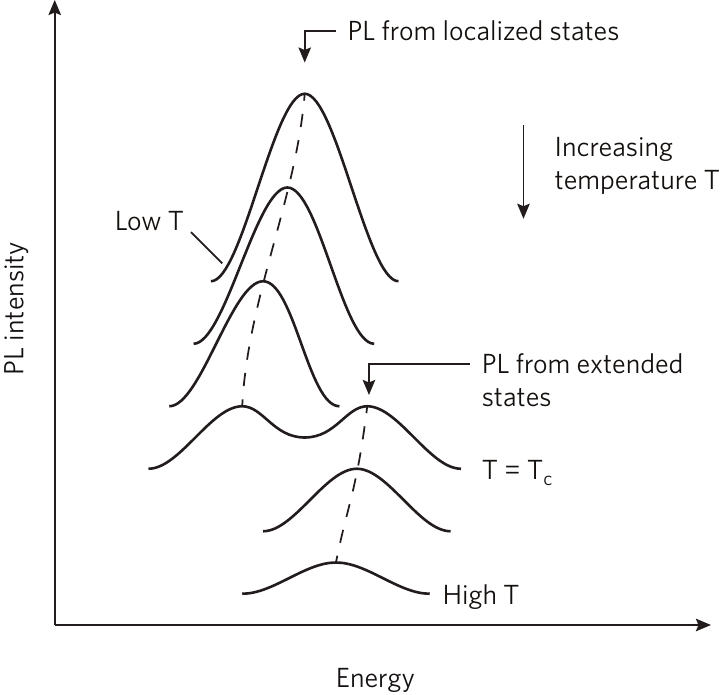}\\
	\caption{PL evolution with increasing temperature (schematic illustration). The low-temperature PL is dominated by emission from localized states, while extended states contribute to the high-temperature spectra. The transition takes place at a critical temperature $T_c$; it is accompanied by widening of the PL spectrum in a narrow temperature region near $T_c$.}\label{Fig:PL-schematic}
\end{figure}

Compound semiconductors and their solid solutions grown at non-equilibrium conditions are susceptible to formation of defect states (deep traps) \cite{Harris_MBE-Growth-Characterization_2005,Batool2013}. Those defects act as non-radiative recombination centers and limit a carrier lifetime, in particular at higher temperatures. This process competes with band-to-band radiative recombination that involves extended as well as shallow localized states (see Fig.~\ref{Fig:PL-schematic}) formed due to composition fluctuations \cite{Ouadjaout_PRB_41_1990}. The non-radiative recombination manifests in quenching of the photoluminescence (PL) yield $\eta$ with increasing temperature $T$ that is accessible experimentally for a wide range of materials, including recent studies of hybrid halide perovskites \cite{Wu_PCCP_16_2014,He_NC_7_2016,Diroll_AFM_27_2017}, 2D materials \cite{Jadczak_N_28_2017}, group III-V dilute bismides \cite{Wilson_SR_8_2018,Hidouri_SM_129_2019}, and GaN:Mg \cite{Reshchikov_PRB_97_2018}. Measurements of $\eta(T)$ provide access to material characteristics---such as a localization energy, density of localized states and traps---when combined with a physics-based theory.

Under continuous-wave (CW) excitation conditions, the carrier generation rate $G$ is balanced by radiative and non-radiative recombination rates
\begin{equation}\label{Eq:G}
  G = R_{r} + R_{nr}.
\end{equation}
The PL yield can be expressed in terms of the non-radiative recombination rate as
\begin{equation}\label{Eq:eta}
  \eta = 1 - R_{nr}/G.
\end{equation}
Shockley, Read, and Hall \cite{Shockley_PR_87_1952,Hall_PR_87_1952} (SRH) proposed the non-radiative recombination rate in the form
\begin{equation}\label{Eq:R_nr-SRH-original}
  R_{nr}=
  \frac
    {
      C_n C_p N_t p n
    }
    {
      C_n n + C_p p
    },
\end{equation}
where $n$ and $p$ are the concentration of electrons and holes in extended states, $C_n$ and $C_p$ are the corresponding capture coefficients, and $N_t$ is the trap density. Here, the thermal generation is assumed to be negligible (see note in Sec.~\ref{Sec:Method}). SRH expression is widely used to describe recombination via traps in device simulations \cite{Li_PRB_95_2017}. Equation~(\ref{Eq:R_nr-SRH-original}) is general and does not require a modification even in the presence of localized states. However, the presence of localized states will greatly influence the carrier concentrations $n(T,G)$ and $p(T,G)$. Finding an exact analytical solution taking into account various competing processes (even without involving localized states) is a formidable task. This inspired a number of approximations and Monte-Carlo simulations in attempt to describe $\eta(T)$ dependence that will be briefly reviewed below.

Starting with \citet{Gee_PRL_42_1979}, a stream of theories emerged where localized states are explicitly considered as the \textit{only} source of radiative recombination. The non-radiative recombination was attributed to a thermal activation of excitons from localized states to the mobility edge. This model was later augmented \cite{Rubel_PRB_73_2006} to include recapture of excitons into radiative states with the probability of $N_l/(N_t+N_l)$, where $N_t$ and $N_l$ correspond to the concentration of non-radiative traps and localized states, respectively. \citet{Shakfa_JAP_117_2015} supplemented the ratio of concentrations with different capture cross sections for traps and localized states. The relative PL yield as a function of temperature becomes \cite{Gee_PRL_42_1979,Rubel_PRB_73_2006,Shakfa_JAP_117_2015}
\begin{equation}\label{Eq:Single-expon-OR-SB}
  \eta =
  \left[
  	1 + \nu_0 \tau_0 \frac{\sigma_t N_t}{\sigma_l N_l + \sigma_t N_t} \exp \left( \frac{-E_l}{k T} \right)
  \right]^{-1}
\end{equation}
for a set of localized states with the localization energy $E_l$. Here, $\nu_0$ is the attempt to escape frequency, $\tau_0$ is the radiative lifetime of excitons captured into localized states, $\sigma$ is the capture cross section which is different for two kind of states, and $k$ is the Boltzmann constant. The advantage of Eq.~(\ref{Eq:Single-expon-OR-SB}) is that it can be readily extended to an arbitrary function for the density of localized states. However, this stream of theories has two deficiencies: (i) excitons are assumed even at temperatures $k T$ that greatly exceed the exciton binding energy, and (ii) extended states are assumed to be dark. The second limitation was lifted by \citet{Jandieri_PRB_87_2013}, but the analysis was presented for $T=0$~K only.

Among empirical dependencies for fitting the temperature quenching of the integrated PL intensity, a double-exponential function is a popular choice \cite{Lambkin_APL_65_1994}
\begin{equation}\label{Eq:Double-expon}
  \eta =
  \left[
    1 +
    C_\text{A} \exp\left(\frac{-E_\text{A}}{k T}\right) +
    C_\text{B} \exp\left(\frac{-E_\text{B}}{k T}\right)
  \right]^{-1}.
\end{equation}
It incorporates two thermally activated non-radiative recombination channels A and B, characterized by activation energies $E_\text{A} < E_\text{B}$. The energy barrier $E_\text{A}$ is typically of the order 10~meV \cite{Sun_APL_89_2006}. It is often ascribed to release of carriers from localized state implying that $E_\text{A} = E_l$ similar to Eq.~(\ref{Eq:Single-expon-OR-SB}). The energy $E_\text{B}$ is of the order of 100~meV, and its interpretation is more broad, including confinement energy for quantum heterostructures \cite{Sun_APL_89_2006} or a reorganization energy associated with vibronic transitions \cite{Stoneham_RPP_44_1981}. The pre-exponential factors (typically, $C_\text{A} \ll C_\text{B}$) are interpreted as a ratio between radiative and non-radiative lifetime for each individual channel \cite{Daly_PRB_52_1995}. This ratio is proportional to the concentration of non-radiative centers \cite{Buyanova_APL_82_2003}. Equation~(\ref{Eq:Double-expon}) performs remarkably well for a variety of materials including chlorine-doped ZnSe layers \cite{Wang_JAP_90_2001}, Ga(AsSbN) quantum wells \cite{Lourenco_JAP_93_2003}, and hybrid halide perovskites \ce{CH3NH3PbBr3} \cite{He_NC_7_2016} to name a few. However, a relation between the double-exponential dependence and the SRH statistics remains obscure.

The goal of this paper is to bridge the gap between the original SRH recombination statistics, which exclusively relies on extended states, and the stream of theories that focus on recombination of localized excitations. The model proposed here extends the well-established SRH recombination statistics and explicitly includes localized states as illustrated in Fig.~\ref{Fig:Proc-schem}. The recombination is treated as a bimolecular process rather than a monomolecular recombination of excitons, as the former is physically more plausible for a wide range of temperatures. It is anticipated that localized states will have a profound influence on the recombination statistics and its temperature dependence. The emphasis is placed on an interplay between two distinct channels of radiative recombination (shallow localized states \textit{vs} extended states) mediated by trapping of photogenerated charge carriers by non-radiative centers.  The final goal is not only to present a numerical solution for a given set of parameters, but also to derive an approximate analytical expression for the thermal quenching of the PL intensity in non-degenerate semiconductors at the limit of low pump intensities. The result is later compared to the double-exponential Eq.~(\ref{Eq:Double-expon}) to discuss its merit.

%
%
\section{Rate equations}\label{Sec:Method}

\begin{figure}[h]
 \includegraphics{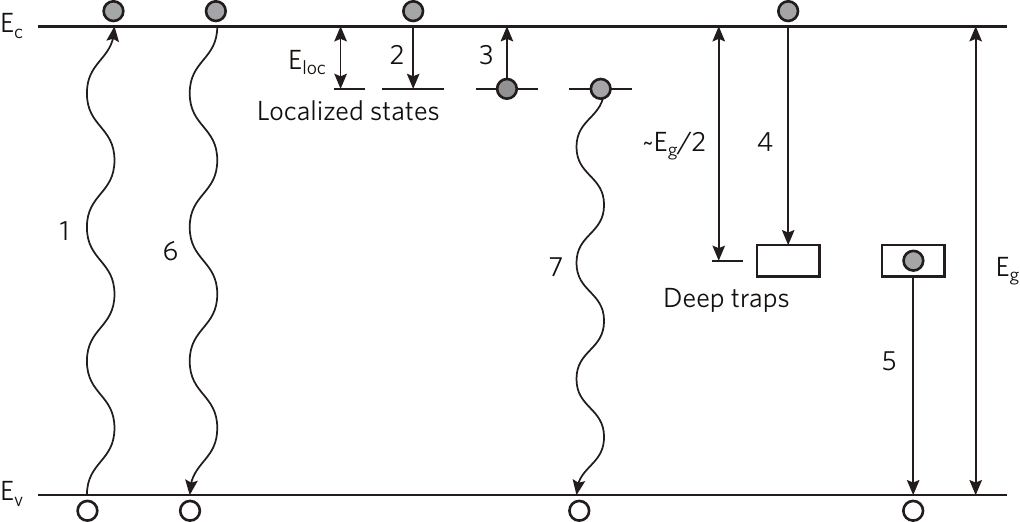}\\
	\caption{Processes included in the model: 1 -- electron-hole pair generation, 2 -- an electron capture to a localized state, 3 -- an electron release to extended states, 4 and 5 -- an electron/hole capture by a deep trap, respectively, 6 and 7 -- radiative recombination of an electron in extended and localized states, respectively. See Table~\ref{Table:1} for the list of symbols.}\label{Fig:Proc-schem}
\end{figure}

Relevant excitation and recombination processes are illustrated in Fig.~\ref{Fig:Proc-schem}. They include transitions originally considered by \citet{Shockley_PR_87_1952} and additional ones due to localized states. All recombination events are treated as bimolecular processes that require the knowledge of concentration of electrons and holes. The total concentration of electrons is subdivided into three components, which represent free electrons at the mobility edge $n_c$, electrons trapped into non-radiative deep centers $n_t$, and those captured into localized states $n_l$. Their time evolution is governed by the set of coupled non-linear rate equations presented below:
\begin{align} \label{Eq:dn/dt}
 \frac{dn_c}{dt}  = &
 - C_n n_c (N_t - n_t)
 - C_n n_c (N_l - n_l) \nonumber\\
 & + C_n n_l N_c \exp \left( \frac{-E_l}{k T} \right)
 - B_{cv} n_c p
 + G.
\end{align}
Here, the order of terms corresponds to the processes 4, 2, 3, 6, and 1 shown in Fig.~\ref{Fig:Proc-schem}. The comprehensive list of model parameters can be found in Table~\ref{Table:1}. The rate equation for electrons in deep traps is
\begin{equation} \label{Eq:dn_t/dt}
 \frac{dn_t}{dt}  =
 C_n n_c (N_t - n_t)
 - C_p n_t p.
\end{equation}
The two terms represent processes 4 and 7 in Fig.~\ref{Fig:Proc-schem}. The rate equation for electrons in localized states is
\begin{align} \label{Eq:dn_l/dt}
 \frac{dn_l}{dt}  = &
 ~C_n n_c (N_l - n_l)
 - C_n n_l N_c \exp \left( \frac{-E_l}{k T} \right) \nonumber\\
 & - B_{lv} n_l p,
\end{align}
where the terms reflect processes 2, 3, and 7 in Fig.~\ref{Fig:Proc-schem}. The concentration of holes is governed by
\begin{equation} \label{Eq:dp/dt}
 \frac{dp}{dt}  =
 - C_p n_t p
 - B_{cv} n_c p
 - B_{lv} n_l p
 + G.
\end{equation}
The order of terms is mapped to processes 5, 6, 7, and 1 in Fig.~\ref{Fig:Proc-schem}. The focus of this paper is on the CW excitation condition, i.e., we are looking for a steady-state solution of Eqs.~(\ref{Eq:dn/dt})--(\ref{Eq:dp/dt}).

It should be noted that the temperature enters into the model not only through the exponential term $\exp(- E_l/k T)$. The capture coefficients are proportional to the thermal velocity of charge carriers leading to $C_n,C_p \propto T^{1/2}$ (Table~\ref{Table:1}). The effective density of extended states also depends on the temperature as $N_c \propto T^{3/2}$. The bimolecular recombination coefficients are kept temperature independent for the sake of simplicity.

The original \citet{Shockley_PR_87_1952} model also includes thermally-assisted release of electrons trapped into deep non-radiative states and capture of electrons from the valence band. These processes are disregarded here for simplicity. This assumption can be justified, provided the following conditions are fulfilled: $E_t \approx E_g/2$, $k T \ll E_t$ and the photo-generation rate $G$ is high enough to keep $n_c p \gg N_c N_p \exp(-E_g/k T)$ at all temperatures under consideration.

%
%
\section{Results and discussion}\label{Sec:Results}

\subsection{Critical temperature $T_c$}

Temperature-dependent PL measurements performed on disordered semiconductors often reveal a so-called ``S-shape" shown schematically in Fig.~\ref{Fig:PL-schematic}. Here, the low-temperature PL is attributed to emission from localized states, while the high-temperature PL is dominated by recombination of free charge carriers. The transition between two regimes occurs at a critical temperature $T_c$. This temperature corresponds to a simultaneous broadening and shift of PL spectrum to higher energies. This effect is also sensitive to the pump intensity $G$ \cite{Wilson_SR_8_2018}, which is related to saturation of localized states at high intensities. The saturation generation rate can be defined as
\begin{equation}\label{Eq:G_0}
  G_0 = B_{lv} N_l^2,
\end{equation}
which represents the upper limit of the radiative recombination rate through localized states under CW excitation. The low excitation intensity conditions are referred to $G/G_0 < 1$.

\begin{figure}[h]
 \includegraphics{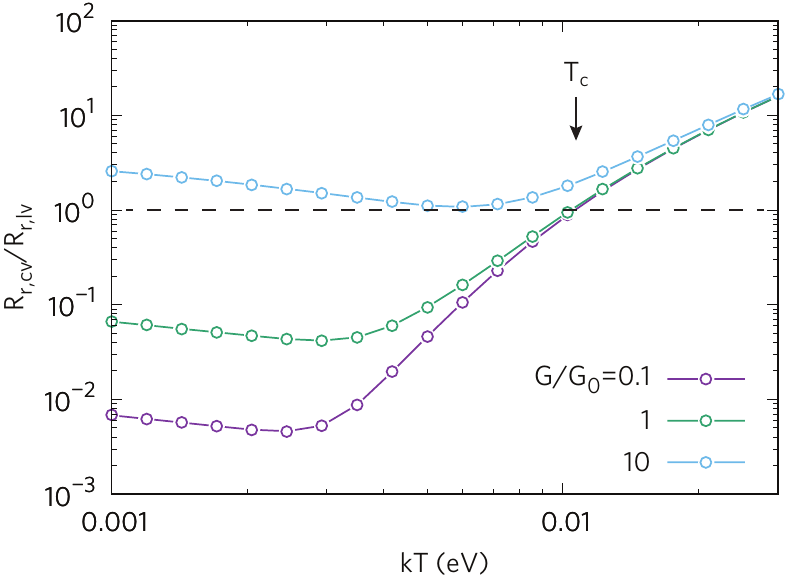}\\
	\caption{Temperature dependence of the free/localized CW radiative recombination ratio at various generation rates related to $G_0$ in Eq.~(\ref{Eq:G_0}).}\label{Fig:LSRH-exact-Rratio-vs-kT_var-G}
\end{figure}

The relative contribution of the two radiative channels ($R_{r,lv}$ for localized states and $R_{r,cv}$ for extended states) to PL as a function of temperature is shown in Fig.~\ref{Fig:LSRH-exact-Rratio-vs-kT_var-G}. Results are presented for three $G/G_0$ ratios to illustrate a difference between low and high excitation intensities. The excitation intensity affects the low-temperature PL composition in favor of recombination from localized states at low intensities. The transition temperature $k T_c \approx 0.01$~eV corresponds to the equal ratio of localized/extended states contribution to PL. We can see that it is \textit{less} than the localization energy $E_l = 0.02$~eV used as the model parameter.

\begin{figure}[h]
 \includegraphics{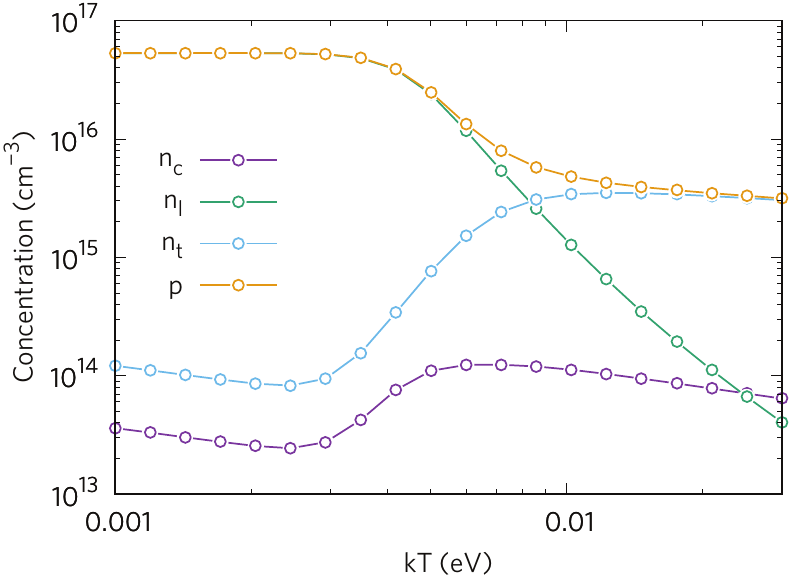}\\
	\caption{Steady-state carrier density as a function of temperature at low excitation intensity [$G/G_0 = 0.1$, see Eq.~(\ref{Eq:G_0})].}\label{Fig:LSRH-exact-conc-vs-kT_G=0.1}
\end{figure}

At the crossover temperature $T_c$ both localized and extended electron-hole pair recombination rates have equal contribution
\begin{equation}\label{Eq:recomb-rate-at-T_c}
  B_{lv} n_l p = B_{cv} n_c p,
\end{equation}
which yields the ratio of electron concentrations
\begin{equation}\label{Eq:loc-state-recomb-balance}
  \frac{n_l}{n_c} = \frac{B_{cv}}{B_{lv}}.
\end{equation}
At steady-state conditions and $G<G_0$ (i.e., $N_l - n_l \approx N_l$), Eq.~(\ref{Eq:dn_l/dt}) yields
\begin{equation}\label{Eq:loc-state-detailed-balance}
  C_n n_c N_l
  - C_n n_l N_c \exp \left( \frac{-E_l}{k T} \right)
  - B_{lv} n_l p = 0.
\end{equation}
Since $T_c$ is high enough, most of the generated carriers recombine non-radiatively. The hole concentration can be approximated as
\begin{equation}\label{Eq:p-high-T}
  p \approx \sqrt{G/C_p}.
\end{equation}
It originates from Eq.~(\ref{Eq:dp/dt}) after neglecting radiative terms and taking into account that the majority of photogenerated electrons reside in traps, implying $n_t = p$ at $T \gtrsim T_c$ (see Fig.~\ref{Fig:LSRH-exact-conc-vs-kT_G=0.1}). After combining Eqs.~(\ref{Eq:loc-state-recomb-balance})--(\ref{Eq:p-high-T}) we obtain
\begin{equation}\label{Eq:T_c}
  T_c = - E_l \left\{ k \ln \left[ \frac{B_{lv}}{N_c} \left( \frac{N_l}{B_{cv}} - \frac{\sqrt{G/C_p}}{C_n} \right) \right] \right\}^{-1}
\end{equation}
It is a transcendental equation since $N_c$, $C_n$, and $C_p$ also depend on temperature (see Table~\ref{Table:1}). The analytical expression yields $k T_c = 0.011$~eV, which agrees with the numerical result in Fig.~\ref{Fig:LSRH-exact-Rratio-vs-kT_var-G}. The dominant term in the square brackets in Eq.~(\ref{Eq:T_c}) is $B_{lv} N_l/B_{cv} N_c=0.15$ at $T_c$. It is responsible for a sizable difference between $k T_c$ and $E_l$. It should be mentioned that qualitatively similar behavior is observed when recombination of localized excitons is considered \cite{Rubel_JAP_98_2005}.

\subsection{Numerical results for $\eta(T)$}

\begin{figure}[h]
 \includegraphics{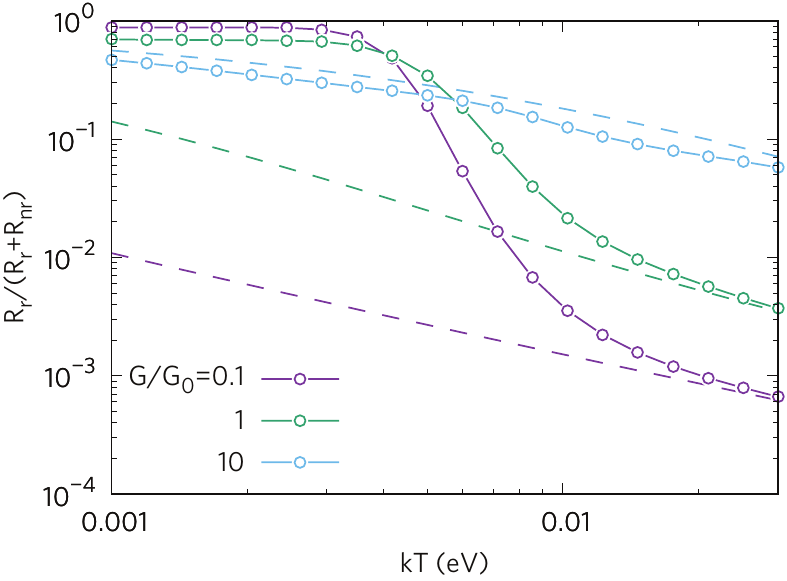}\\
	\caption{Temperature dependence of the CW radiative recombination efficiency at various generation rates related to $G_0$ in Eq.~(\ref{Eq:G_0}). Circles show the model with localized states. Dashed lines correspond to the original SRH model without localized states.}\label{Fig:LSRH-exact-REff-vs-kT_var-G}
\end{figure}

The radiative recombination efficiency is shown in Fig.~\ref{Fig:LSRH-exact-REff-vs-kT_var-G} as a function of temperature at several generation rates. The generation rates are selected such as to cover the range of high rates ($G/G_0 \gg 1$, saturated localized states), low rates ($G/G_0 \ll 1$, negligible population effects), and an intermediate condition. The thermal quenching of PL is most pronounced at low generation rates; it drops by 3 orders of magnitude when approaching the room temperature. The PL efficiency shows a characteristic plateau at low temperatures that is often observed experimentally. Previously, it was attributed to a hopping energy relaxation of recombining excitations \cite{Rubel_PRB_73_2006}. Current results suggest that the same effect can also be observed in a multiple trapping regime even with a monoenergetic distribution of localized states. It is intriguing to decompose the PL quenching into contributions from localized and extended states.

Dashed lines in Fig.~\ref{Fig:LSRH-exact-REff-vs-kT_var-G} correspond to the same material parameters (density of traps, etc.) with processes 2, 3, and 7 (Fig.~\ref{Fig:Proc-schem}) eliminated from the model. The remaining processes are those limited to extended states and traps considered originally by \citet{Shockley_PR_87_1952}. The PL yield without localized states demonstrates a \textit{non-exponential} (power) temperature dependence
\begin{equation}\label{Eq:eta_SRH}
  \eta_\text{SRH} \propto T^{\alpha}
\end{equation}
with $\alpha = -1.1 \ldots -0.6$ for different generation rates $G$. Its origin will be discussed later.

Interestingly, the model with localized states and the original SRH model agree in the limit of high temperatures or high excitation intensities. Neither  Eq.~(\ref{Eq:Single-expon-OR-SB}) nor Eq.~(\ref{Eq:Double-expon}) allude to this result. The PL yield $\eta$ in the presence of localized states is substantially higher than in the case of free carrier recombination. Thus, localized states promote the radiative recombination, particularly at low temperatures. This conclusion is not apparent in the framework of the excitonic model \cite{Rubel_PRB_73_2006,Shakfa_JAP_117_2015}, where only one source of radiative recombination (localized states) is considered.

At first glance, this result seems contradicting experimental observation. Usually, crystalline semiconductors with a strongest localization (e.g., dilute nitrides and bismides) also exhibit greatest PL thermal quenching. The key factor here is growth conditions \cite{Borkovska_TSF_515_2006}. Highly mismatched alloys are grown at non-equilibrium growth conditions that favor incorporation of nitrogen or bismuth. Their growth is also accompanied by a higher defect/trap density $N_t$ at the same time.

\subsection{Analytical approximation for $\eta(T)$}

The nonlinear set of Eqs.~(\ref{Eq:dn/dt})--(\ref{Eq:dp/dt}) is prohibitively complicated for a general solution even in the steady-state regime. Our strategy is to derive analytical approximation for two limits ($T=0$~K and $T>T_c$) and combine them together. We will aim at $G \ll G_0$ as it is the most interesting regime for studding traces of localization. Also, we can neglect finite population effects in this regime, assuming that the carrier density is much less than the density of states (including traps). The starting point is Eq.~(\ref{Eq:eta}) with the non-radiative rate $R_{nr}$ approximated as [see Eq.~(\ref{Eq:dn_t/dt})]
\begin{equation}\label{Eq:R_nr-approx}
  R_{nr} \approx C_n n_c N_t
\end{equation}
with the unknown temperature-dependent concentration of free electrons $n_c$.

In the low-temperature limit radiative recombination of free carriers and thermal excitation of electrons captured in localized states can be neglected. Under such circumstances, Eq.~(\ref{Eq:dn/dt}) takes the form
\begin{equation}\label{Eq:dn/dt-low-T}
  G = C_n n_c N_l + C_n n_c N_t .
\end{equation}
It leads to the free electron concentration
\begin{equation}\label{Eq:n_c-low-T}
  n_c(T=0) = \frac{G}{C_n (N_l + N_t)}.
\end{equation}
After combining Eqs.~(\ref{Eq:eta}), (\ref{Eq:R_nr-approx}), and (\ref{Eq:n_c-low-T}) we obtain the radiative efficiency in the low-temperature limit
\begin{equation}\label{Eq:eta-low-T}
  \eta(T=0) \approx \frac{N_l}{N_l + N_t}.
\end{equation}

In the high-temperature limit ($T > T_c$), it is convenient to present the steady-state limit of Eq.~(\ref{Eq:dn/dt}) using Eq.~(\ref{Eq:dn_l/dt}) as
\begin{equation}\label{Eq:dn/dt-high-T}
  G = B_{lv} n_l p + C_n n_c N_t + B_{cv} n_c p
\end{equation}
Here, the balance of capture and release from localized states is replaced by the recombination rate from localized states. Even though the term $B_{lv} n_l p$ in Eq.~(\ref{Eq:dn/dt-high-T}) is not the leading term, it is kept for the sake of comparison of the result with the low-temperature limit at a later stage. At high temperatures localized and extended states have equal fractional occupancy leading to
\begin{equation}\label{Eq:n_l/n_c-high-T}
  \frac{n_l}{n_c} = \frac{N_l}{N_c}.
\end{equation}
Combining this result and Eq.~(\ref{Eq:p-high-T}), the high-temperature radiative rate from localized states becomes
\begin{equation}\label{Eq:R_r,lv-high-T}
  B_{lv} n_l p = B_{lv} n_c \frac{N_l}{N_c} \sqrt{G/C_p}.
\end{equation}
The equilibrium concentration of free electrons can be found after combining Eqs.~(\ref{Eq:p-high-T}), (\ref{Eq:dn/dt-high-T}), and (\ref{Eq:R_r,lv-high-T})
\begin{equation}\label{Eq:n_c-high-T}
  n_c(T > T_c) = \frac{G}{C_n N_t + B_{lv} (N_l/N_c) \sqrt{G/C_p} +  B_{cv}\sqrt{G/C_p}}.
\end{equation}
After combining Eqs.~(\ref{Eq:eta}), (\ref{Eq:R_nr-approx}), and (\ref{Eq:n_c-high-T}) we obtain the radiative efficiency in the high-temperature limit
\begin{widetext}
  \begin{equation}\label{Eq:eta-high-T}
    \eta(T > T_c) \approx 1 -
    \left(
      1 + \frac{B_{lv} N_l}{N_t N_c C_n} \sqrt{G/C_p} + \frac{B_{cv}}{C_n N_t} \sqrt{G/C_p}
    \right)^{-1}.
  \end{equation}
\end{widetext}
The obtained result has a general form of
\begin{equation}\label{Eq:eta-ratios}
  \eta \approx 1 - (1 + u_{lv} + u_{cv})^{-1},
\end{equation}
where $u$ is a ratio between a radiative recombination rate for a specific channel and the non-radiative recombination rate
\begin{equation}\label{Eq:u-def}
  u_{lv(cv)} = \frac{R_{r,lv(cv)}}{R_{nr}}.
\end{equation}
Since the free carrier recombination dominates at high temperatures and low generation rates (Fig.~\ref{Fig:LSRH-exact-Rratio-vs-kT_var-G}), Eq.~(\ref{Eq:eta-high-T}) can be further simplified
\begin{equation}\label{Eq:eta-high-T-approx}
  \eta(T > T_c) \approx \frac{B_{cv}}{C_n N_t} \, \sqrt{G/C_p}.
\end{equation}

Generalization of $\eta$ to an arbitrary $T$ requires a better approximation for $u_{lv}$. Equations~(\ref{Eq:eta-low-T}) and (\ref{Eq:eta-high-T}) infer the low- and high-temperature limits
\begin{equation}\label{Eq:u_lv-two-limits}
  u_{lv} =
    \begin{cases}
      N_l/N_t       & \quad \text{if } T=0, \\
      (N_l/N_t) \, (B_{lv}/N_c C_n) \sqrt{G/C_p}  & \quad \text{if } T > T_c .
    \end{cases}
\end{equation}
Since the carrier release from localized states is an activated process, we can employ an exponential sigmoidal function
\begin{equation}\label{Eq:u_lv-any-T}
  u_{lv}(T) \approx \frac{N_l/N_t}{1 + (N_c C_n/B_{lv}) \sqrt{C_p/G} \exp(-E_l/k T)}
\end{equation}
to approximate the $T$ dependence with proper limits and the activation energy $E_l$.

After substitution of Eq.~(\ref{Eq:u_lv-any-T}) into Eq.~(\ref{Eq:eta-ratios}) the expression for the temperature-dependent PL intensity reads
\begin{widetext}
  \begin{equation}\label{Eq:Eff-approx}
    \eta(T) \approx
    1 -
    \left\{
      1 +
      \frac{N_l}{N_t}\left[ 1 + A_1^\circ \left( \frac{T}{T^\circ} \right)^{\alpha_1} \exp \left( \frac{-E_l}{k T} \right) \right]^{-1} +
      u_{cv}^\circ \left( \frac{T}{T^\circ} \right)^{\alpha_2}
    \right\}^{-1}.
  \end{equation}
\end{widetext}
Here, the degree symbol ($^\circ$) indicates a value of material parameters taken at standard conditions. The following dimensionless parameters are used
\begin{equation}\label{Eq:A_1}
  A_1 = \frac{N_c C_n}{B_{lv}} \sqrt{\frac{C_p}{G}}
\end{equation}
and
\begin{equation}\label{Eq:u_cv}
  u_{cv} = \frac{B_{cv}}{C_n N_t} \, \sqrt{G/C_p} .
\end{equation}
The exponents
\begin{equation}\label{Eq:alpha_1}
  \alpha_1 = 3/2 + 1/2 + 1/4 = 9/4
\end{equation}
and
\begin{equation}\label{Eq:alpha_2}
  \alpha_2 = -1/2 - 1/4 = -3/4
\end{equation}
are governed by the temperature dependence of $C_p$, $C_n$, and $N_c$ (Table~\ref{Table:1}). The expectation value for $\alpha_2$ agrees with the exponent in Eq.~(\ref{Eq:eta_SRH}) obtained by fitting to numerical results.

\begin{figure}[h]
 \includegraphics{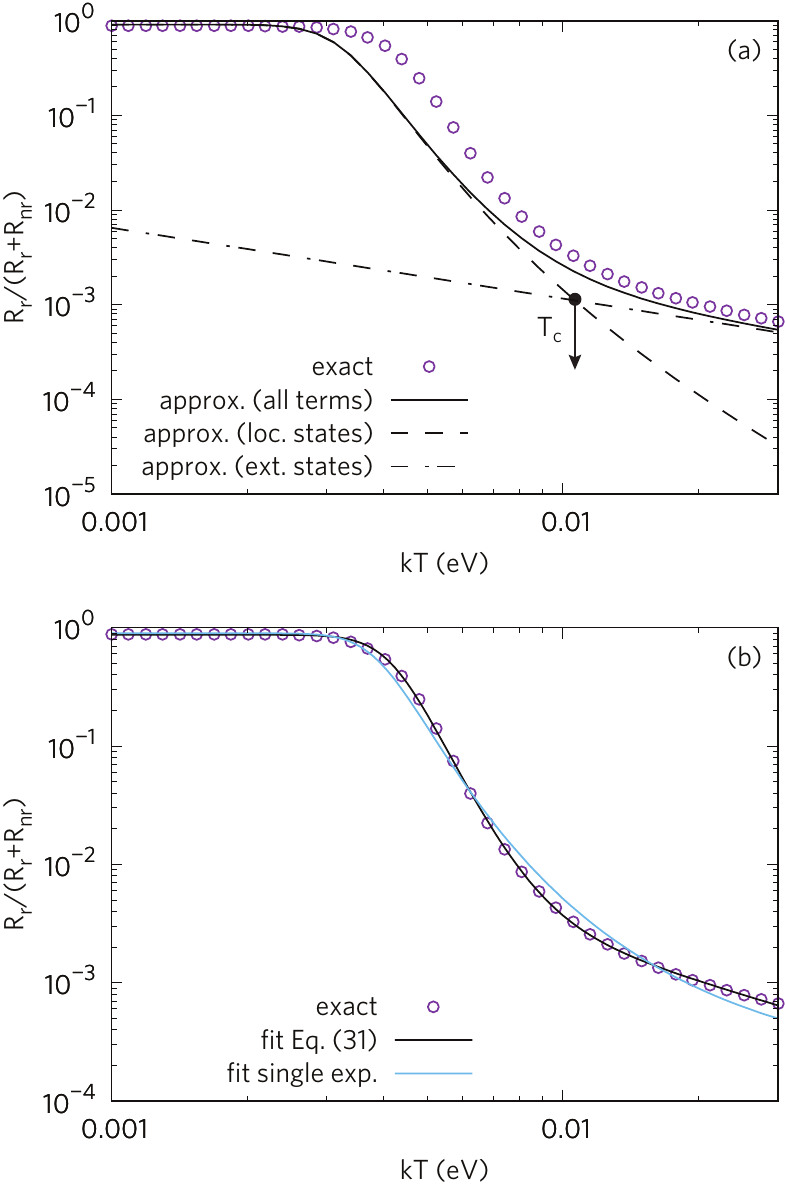}\\
	\caption{Temperature dependence of the CW radiative recombination efficiency calculated using different methods: (a) an exact numerical solution of Eqs.~(\ref{Eq:dn/dt})--(\ref{Eq:dn_l/dt}) and the approximate analytical Eq.~(\ref{Eq:Eff-approx}) with the model parameters in Table~\ref{Table:1} and the generation rate $G = G_0/10$, (b) fitting to numerical results using the single exponential form of Eq.~(\ref{Eq:Double-expon}) with parameters $C_\text{A} = 5807$ and $E_\text{A} = 0.035$~eV \textit{vs} fitting with Eq.~(\ref{Eq:Eff-approx}) using parameters $N_l/N_t = 6.6$, $A_1^\circ = 2.0 \times 10^6$, $kT^\circ = 0.026$~eV, $\alpha_1 = 4.2$, $E_l = 0.021$~eV, $u_{cv}^\circ = 7.5 \times 10^{-4}$, and $\alpha_2 = -1.1$. The approximate analytical solution on panel (a) is broken into two contributions: recombination from localized states and extended states. The crossover between the two mechanisms takes place at $T_c$.}\label{Fig:LSRH-approx-fit-REff-vs-kT_G=0.1}
\end{figure}

Figure~\ref{Fig:LSRH-approx-fit-REff-vs-kT_G=0.1}(a) illustrates a comparison between the exact numerical solution and Eq.~(\ref{Eq:Eff-approx}) with model parameters from Table~\ref{Table:1}. It should be stressed that the solid line in Figure~\ref{Fig:LSRH-approx-fit-REff-vs-kT_G=0.1}(a) is not a fitting to the data points. We can see that $\eta(T)$ is governed by recombination of localized carriers at $T < T_c$ and free carriers at $T > T_c$. The agreement between the approximate analytical solution and numerical results is satisfactory.

Performance of a single-exponential form of Eq.~(\ref{Eq:Double-expon}) is evaluated by fitting to the exact numerical results in Fig.~\ref{Fig:LSRH-approx-fit-REff-vs-kT_G=0.1}(b). The single-exponential form is selected due to the presence of only one non-radiative channel in our model. The success or failure of the fitting function is judged based on its ability to extract meaningful material parameters rather than a minimum of the residual error on the plot. The extracted fitting parameters are $C_\text{A} = 5.8 \times 10^3$ and $E_\text{A} = 0.035$~eV. The pre-exponential factor is the radiative/non-radiative lifetime ratio in the high-$T$ limit, which corresponds to the inverse ratio or rates $(u_{cv}^\circ)^{-1} = 2.0 \times 10^3$ that agrees reasonably well with $C_\text{A}$. This factor is indeed proportional to the trap density $N_t$ as can be seen in Eq.~(\ref{Eq:u_cv}). The apparent activation energy $E_\text{A}$ overestimates the localization energy $E_l = 0.02$~eV almost twice.

The fitting performance of Eq.~(\ref{Eq:Eff-approx}) was also tested and shown in Fig.~\ref{Fig:LSRH-approx-fit-REff-vs-kT_G=0.1}(b). The fit yields $u_{cv}^\circ = 7.5 \times 10^{-4}$ \textit{vs} the anticipated value of $5.1 \times 10^{-4}$, $E_l = 0.021$~eV \textit{vs} 0.02~eV, and $\alpha_2 = -1.1$ \textit{vs} $-3/4$. Even though Eq.~(\ref{Eq:Eff-approx}) is more complex, it provides more reliable parameters that the single-exponential Eq.~(\ref{Eq:Double-expon}).

\subsection{Proportionality between generation and radiative recombination rates}

\citet{Mazur_JAP_113_2013} noted that the PL intensity in GaAs$_{1-x}$Bi$_x$/GaAs quantum well is linearly proportional to excitation intensity $G$ at a low temperature and low $G$; the dependence becomes weaker at higher $G$'s. On the other hand, a superliner dependence of the PL intensity \textit{vs} $G$ was reported for \ce{CH3NH3PbBr3} films with the exponent of 1.2 \cite{He_NC_7_2016}. The exponent also tends to increase with temperature for (GaIn)(SbBi) and Ga(AsBi) layers \cite{Linhart_JPDAP_50_2017,Wilson_SR_8_2018}.

\begin{figure}[h]
 \includegraphics{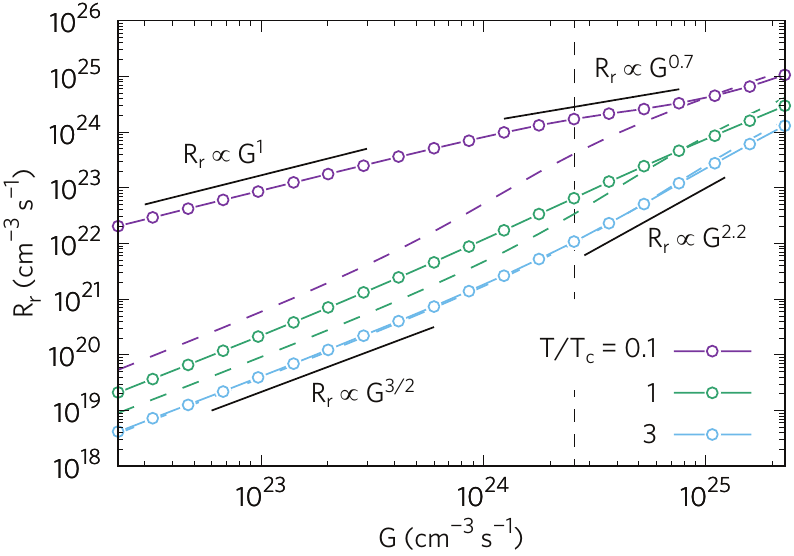}\\
	\caption{CW radiative recombination rate as a function of generation at various temperatures. Circles show the model with localized states. Dashed lines correspond to the original SRH model without localized states. The vertical line is the saturation generation rate $G_0$ defined in Eq.~(\ref{Eq:G_0}). The temperature is measured relative to the critical temperature $T_c$ [solution of Eq.~(\ref{Eq:T_c})].}\label{Fig:LSRH-exact-R-vs-G_var-kT}
\end{figure}

Results of calculations obtained in the framework of the present model (Fig.~\ref{Fig:LSRH-exact-R-vs-G_var-kT}) suggest a general relationship between excitation and recombination rates in the form
\begin{equation}\label{Eq:R_r-proportionality}
  R_r \propto G^\beta.
\end{equation}
The exponent $\beta = 0.7 - 2.2$ is sensitive to the presence of localized states in the model and also varies with the temperature and excitation intensity very similar to the experimental report by \citet{Wilson_SR_8_2018}. It approaches the value of $\beta = 1$ at low $G$ and $T$ \textit{only} when localized states dominate in the PL emission (purple circles in Fig.~\ref{Fig:LSRH-exact-R-vs-G_var-kT}). The original SRH model (without localized states) has a different exponent $\beta \approx 3/2$ under identical conditions (purple dashed line in Fig.~\ref{Fig:LSRH-exact-R-vs-G_var-kT}). The same exponent ($\beta \approx 3/2$) is observed in the model with localized states at high temperatures $T > T_c$ and low generation rates $G < G_0$.

Next, we consider few limiting cases to elucidate the origin of the exponent $\beta$ in Eq.~(\ref{Eq:R_r-proportionality}). The generation and recombination rates are linked \textit{via}
\begin{equation}\label{Eq:R_r-general}
  R_r = \eta \, G ,
\end{equation}
that results from Eqs.~(\ref{Eq:G}) and (\ref{Eq:eta}). Thus, the linear proportionality between generation and recombination ($\beta = 1$) is only possible when the PL yield $\eta$ is independent of the excitation intensity. It is the case at $T \ll T_c$ and low $G < G_0$ when localized states are present. The PL efficiency is then governed by Eq.~(\ref{Eq:eta-low-T}) that yields the proportionality
\begin{equation}\label{Eq:R_r-low-T}
  R_r(T \ll T_c) \approx \frac{N_l}{N_l + N_t} \, G .
\end{equation}

At higher temperatures ($T > T_c$) the PL yield $\eta$ becomes $G$ dependent as approximated by Eq.~(\ref{Eq:eta-high-T-approx}). As a result, the recombination rate becomes a superlinear function of generation
\begin{equation}\label{Eq:R_r-high-T}
  R_r(T > T_c) \approx \frac{B_{cv}}{C_n N_t \sqrt{C_p}} \, G^{3/2}.
\end{equation}
This result is different from $\beta=0.5$ and $2$ suggested by \cite{Schmidt_PRB_45_1992} for free-to-bound exciton transitions and free carrier recombination, respectively.

The sublinear dependence ($\beta < 1$ as observed in Ref.~\citenum{Mazur_JAP_113_2013}) occurs only in the model with localized states at $T \ll T_c$ and $G \sim G_0$ (Fig.~\ref{Fig:LSRH-exact-R-vs-G_var-kT}). The generation rate is governed by the excitation energy flux. The order of magnitude of $G$ for solar cell applications can be estimated from a product of the photon flux $\Phi$ and the absorption coefficient. Taking the photon flux of $\Phi=3\times10^{17}~\text{cm}^{-2}~\text{s}^{-1}$, which corresponds to the energy flux of $1000~\text{W~m}^{-2}$ at the average photon energy of 2~eV for $\times 1$ sun irradiation, and the absorption coefficient of $10^4~\text{cm}^{-1}$, we obtain $G \sim 3\times10^{21}~\text{cm}^{-3}~\text{s}^{-1}$. This value corresponds to $10^{-3} G_0$ for the set of material parameters in Table~\ref{Table:1}. Thus, it is possible to reach saturation of localized states under $\times 1000$ concentrated sun light, which is realistic under operating conditions for multi-junction solar cells.

%
%
\section{Conclusions}\label{Sec:Conclusions}

The model proposed here extends the well-established Shockley-Read-Hall recombination statistics and explicitly includes localized states. The recombination is treated as a bimolecular process as it is physically more plausible for a wide range of temperatures. The band-to-band radiative recombination involves extended as well as shallow localized states. The low-temperature photoluminescence (PL) is attributed to emission from localized states, while the high-temperature PL is dominated by recombination of free charge carriers. The transition temperature corresponding to the equal ratio of localized/extended states is linked to the localization energy $E_l$, however, this temperature is significantly less than expected from $E_l/k$. It is a relatively high effective density of extended states in the conduction band that promotes thermal release of electrons captured into localized states.

The original Shockley-Read-Hall model (i.e., without localized states) demonstrates a non-exponential (power) temperature dependence of the PL yield. In the presence of localized states, the thermal quenching of PL is most pronounced at low excitation rates when the localized states are not fully saturated. Both models (with localized states and the original Shockley-Read-Hall model) agree in the limit of high temperatures or high excitation intensities when the extended states contribute most to the PL. Localized states promote radiative recombination at low temperatures and enhance the radiative recombination efficiency.

An analytical expression is derived for the temperature-dependent PL yield showing the interplay between two radiative recombination channels. It is compared to a widely-used double-exponential empirical dependence for the thermal quenching of the relative PL intensity. Physical merit and ability to extract material parameters, from fitting to exact numerical solution is tested. The newly proposed expression provides more reliable material parameters, especially the localization energy. Additional insight is given to a pre-exponential factor used in the empirical expression. It is linked not only to the trap density, but also to electron and hole capture coefficients, a bimolecular recombination coefficient for extended states and the excitation rate.

Finally, a relationship between excitation and radiative recombination rates is established. The radiative recombination is linearly proportional to excitation only at low temperatures and low generation rates when localized states dominate in the PL emission. The radiative recombination of excitations from extended states shows a superlinear dependence on the generation rate with the exponent of 3/2.

%
%
\begin{acknowledgments}
Funding provided by the Natural Sciences and Engineering Research Council of Canada
under the Discovery Grant Programs RGPIN-2015-04518 is gratefully acknowledged.
\end{acknowledgments}

%
%

\begin{thebibliography}{29}%
\makeatletter
\providecommand \@ifxundefined [1]{%
 \@ifx{#1\undefined}
}%
\providecommand \@ifnum [1]{%
 \ifnum #1\expandafter \@firstoftwo
 \else \expandafter \@secondoftwo
 \fi
}%
\providecommand \@ifx [1]{%
 \ifx #1\expandafter \@firstoftwo
 \else \expandafter \@secondoftwo
 \fi
}%
\providecommand \natexlab [1]{#1}%
\providecommand \enquote  [1]{``#1''}%
\providecommand \bibnamefont  [1]{#1}%
\providecommand \bibfnamefont [1]{#1}%
\providecommand \citenamefont [1]{#1}%
\providecommand \href@noop [0]{\@secondoftwo}%
\providecommand \href [0]{\begingroup \@sanitize@url \@href}%
\providecommand \@href[1]{\@@startlink{#1}\@@href}%
\providecommand \@@href[1]{\endgroup#1\@@endlink}%
\providecommand \@sanitize@url [0]{\catcode `\\12\catcode `\$12\catcode
  `\&12\catcode `\#12\catcode `\^12\catcode `\_12\catcode `\%12\relax}%
\providecommand \@@startlink[1]{}%
\providecommand \@@endlink[0]{}%
\providecommand \url  [0]{\begingroup\@sanitize@url \@url }%
\providecommand \@url [1]{\endgroup\@href {#1}{\urlprefix }}%
\providecommand \urlprefix  [0]{URL }%
\providecommand \Eprint [0]{\href }%
\providecommand \doibase [0]{http://dx.doi.org/}%
\providecommand \selectlanguage [0]{\@gobble}%
\providecommand \bibinfo  [0]{\@secondoftwo}%
\providecommand \bibfield  [0]{\@secondoftwo}%
\providecommand \translation [1]{[#1]}%
\providecommand \BibitemOpen [0]{}%
\providecommand \bibitemStop [0]{}%
\providecommand \bibitemNoStop [0]{.\EOS\space}%
\providecommand \EOS [0]{\spacefactor3000\relax}%
\providecommand \BibitemShut  [1]{\csname bibitem#1\endcsname}%
\let\auto@bib@innerbib\@empty
\bibitem [{\citenamefont {{Harris Jr.}}\ \emph {et~al.}(2005)\citenamefont
  {{Harris Jr.}}, \citenamefont {Yuen}, \citenamefont {Bank}, \citenamefont
  {Wistey}, \citenamefont {Lordi}, \citenamefont {Gugov}, \citenamefont {Bae},\
  and\ \citenamefont {Goddard}}]{Harris_MBE-Growth-Characterization_2005}%
  \BibitemOpen
  \bibfield  {author} {\bibinfo {author} {\bibfnamefont {J.~S.}\ \bibnamefont
  {{Harris Jr.}}}, \bibinfo {author} {\bibfnamefont {H.}~\bibnamefont {Yuen}},
  \bibinfo {author} {\bibfnamefont {S.}~\bibnamefont {Bank}}, \bibinfo {author}
  {\bibfnamefont {M.}~\bibnamefont {Wistey}}, \bibinfo {author} {\bibfnamefont
  {V.}~\bibnamefont {Lordi}}, \bibinfo {author} {\bibfnamefont
  {T.}~\bibnamefont {Gugov}}, \bibinfo {author} {\bibfnamefont
  {H.}~\bibnamefont {Bae}}, \ and\ \bibinfo {author} {\bibfnamefont
  {L.}~\bibnamefont {Goddard}},\ }\enquote {\bibinfo {title} {Dilute nitride
  semiconductors},}\ \ (\bibinfo  {publisher} {Elsevier Ltd.},\ \bibinfo {year}
  {2005})\ Chap.\ \bibinfo {chapter} {MBE Growth and Characterization of Long
  Wavelength Dilute Nitride {III-V} Alloys}, pp.\ \bibinfo {pages}
  {1--92}\BibitemShut {NoStop}%
\bibitem [{\citenamefont {Batool}\ \emph {et~al.}(2013)\citenamefont {Batool},
  \citenamefont {Chatterjee}, \citenamefont {Chernikov}, \citenamefont {Duzik},
  \citenamefont {Fritz}, \citenamefont {Gogineni}, \citenamefont {Hild},
  \citenamefont {Hosea}, \citenamefont {Imhof}, \citenamefont {Johnson},
  \citenamefont {Jiang}, \citenamefont {Jin}, \citenamefont {Koch},
  \citenamefont {Koch}, \citenamefont {Kolata}, \citenamefont {Lewis},
  \citenamefont {Lu}, \citenamefont {Masnadi-Shirazi}, \citenamefont
  {Millunchick}, \citenamefont {Mooney}, \citenamefont {Riordan}, \citenamefont
  {Rubel}, \citenamefont {Sweeney}, \citenamefont {Thomas}, \citenamefont
  {Thr\"{a}nhardt}, \citenamefont {Tiedje},\ and\ \citenamefont
  {Volz}}]{Batool2013}%
  \BibitemOpen
  \bibfield  {author} {\bibinfo {author} {\bibfnamefont {Z.}~\bibnamefont
  {Batool}}, \bibinfo {author} {\bibfnamefont {S.}~\bibnamefont {Chatterjee}},
  \bibinfo {author} {\bibfnamefont {A.}~\bibnamefont {Chernikov}}, \bibinfo
  {author} {\bibfnamefont {A.}~\bibnamefont {Duzik}}, \bibinfo {author}
  {\bibfnamefont {R.}~\bibnamefont {Fritz}}, \bibinfo {author} {\bibfnamefont
  {C.}~\bibnamefont {Gogineni}}, \bibinfo {author} {\bibfnamefont
  {K.}~\bibnamefont {Hild}}, \bibinfo {author} {\bibfnamefont {T.~J.~C.}\
  \bibnamefont {Hosea}}, \bibinfo {author} {\bibfnamefont {S.}~\bibnamefont
  {Imhof}}, \bibinfo {author} {\bibfnamefont {S.~R.}\ \bibnamefont {Johnson}},
  \bibinfo {author} {\bibfnamefont {Z.}~\bibnamefont {Jiang}}, \bibinfo
  {author} {\bibfnamefont {S.}~\bibnamefont {Jin}}, \bibinfo {author}
  {\bibfnamefont {M.}~\bibnamefont {Koch}}, \bibinfo {author} {\bibfnamefont
  {S.~W.}\ \bibnamefont {Koch}}, \bibinfo {author} {\bibfnamefont
  {K.}~\bibnamefont {Kolata}}, \bibinfo {author} {\bibfnamefont {R.~B.}\
  \bibnamefont {Lewis}}, \bibinfo {author} {\bibfnamefont {X.}~\bibnamefont
  {Lu}}, \bibinfo {author} {\bibfnamefont {M.}~\bibnamefont {Masnadi-Shirazi}},
  \bibinfo {author} {\bibfnamefont {J.~M.}\ \bibnamefont {Millunchick}},
  \bibinfo {author} {\bibfnamefont {P.~M.}\ \bibnamefont {Mooney}}, \bibinfo
  {author} {\bibfnamefont {N.~A.}\ \bibnamefont {Riordan}}, \bibinfo {author}
  {\bibfnamefont {O.}~\bibnamefont {Rubel}}, \bibinfo {author} {\bibfnamefont
  {S.~J.}\ \bibnamefont {Sweeney}}, \bibinfo {author} {\bibfnamefont {J.~C.}\
  \bibnamefont {Thomas}}, \bibinfo {author} {\bibfnamefont {A.}~\bibnamefont
  {Thr\"{a}nhardt}}, \bibinfo {author} {\bibfnamefont {T.}~\bibnamefont
  {Tiedje}}, \ and\ \bibinfo {author} {\bibfnamefont {K.}~\bibnamefont
  {Volz}},\ }in\ \href {\doibase 10.1016/b978-0-12-387839-7.00007-5} {\emph
  {\bibinfo {booktitle} {Molecular Beam Epitaxy}}},\ \bibinfo {editor} {edited
  by\ \bibinfo {editor} {\bibfnamefont {M.}~\bibnamefont {Henini}}}\ (\bibinfo
  {publisher} {Elsevier},\ \bibinfo {year} {2013})\ pp.\ \bibinfo {pages}
  {139--158}\BibitemShut {NoStop}%
\bibitem [{\citenamefont {Ouadjaout}\ and\ \citenamefont
  {Marfaing}(1990)}]{Ouadjaout_PRB_41_1990}%
  \BibitemOpen
  \bibfield  {author} {\bibinfo {author} {\bibfnamefont {D.}~\bibnamefont
  {Ouadjaout}}\ and\ \bibinfo {author} {\bibfnamefont {Y.}~\bibnamefont
  {Marfaing}},\ }\href {\doibase 10.1103/PhysRevB.41.12096} {\bibfield
  {journal} {\bibinfo  {journal} {Phys. Rev. B}\ }\textbf {\bibinfo {volume}
  {41}},\ \bibinfo {pages} {12096} (\bibinfo {year} {1990})}\BibitemShut
  {NoStop}%
\bibitem [{\citenamefont {Wu}\ \emph {et~al.}(2014)\citenamefont {Wu},
  \citenamefont {Bera}, \citenamefont {Ma}, \citenamefont {Du}, \citenamefont
  {Yang}, \citenamefont {Li},\ and\ \citenamefont {Wu}}]{Wu_PCCP_16_2014}%
  \BibitemOpen
  \bibfield  {author} {\bibinfo {author} {\bibfnamefont {K.}~\bibnamefont
  {Wu}}, \bibinfo {author} {\bibfnamefont {A.}~\bibnamefont {Bera}}, \bibinfo
  {author} {\bibfnamefont {C.}~\bibnamefont {Ma}}, \bibinfo {author}
  {\bibfnamefont {Y.}~\bibnamefont {Du}}, \bibinfo {author} {\bibfnamefont
  {Y.}~\bibnamefont {Yang}}, \bibinfo {author} {\bibfnamefont {L.}~\bibnamefont
  {Li}}, \ and\ \bibinfo {author} {\bibfnamefont {T.}~\bibnamefont {Wu}},\
  }\href {\doibase 10.1039/c4cp03573a} {\bibfield  {journal} {\bibinfo
  {journal} {Phys. Chem. Chem. Phys.}\ }\textbf {\bibinfo {volume} {16}},\
  \bibinfo {pages} {22476} (\bibinfo {year} {2014})}\BibitemShut {NoStop}%
\bibitem [{\citenamefont {He}\ \emph {et~al.}(2016)\citenamefont {He},
  \citenamefont {Yu}, \citenamefont {Li}, \citenamefont {Li}, \citenamefont
  {Si}, \citenamefont {Jin}, \citenamefont {Wang}, \citenamefont {Wang},
  \citenamefont {He}, \citenamefont {Wang}, \citenamefont {Zhang},\ and\
  \citenamefont {Ye}}]{He_NC_7_2016}%
  \BibitemOpen
  \bibfield  {author} {\bibinfo {author} {\bibfnamefont {H.}~\bibnamefont
  {He}}, \bibinfo {author} {\bibfnamefont {Q.}~\bibnamefont {Yu}}, \bibinfo
  {author} {\bibfnamefont {H.}~\bibnamefont {Li}}, \bibinfo {author}
  {\bibfnamefont {J.}~\bibnamefont {Li}}, \bibinfo {author} {\bibfnamefont
  {J.}~\bibnamefont {Si}}, \bibinfo {author} {\bibfnamefont {Y.}~\bibnamefont
  {Jin}}, \bibinfo {author} {\bibfnamefont {N.}~\bibnamefont {Wang}}, \bibinfo
  {author} {\bibfnamefont {J.}~\bibnamefont {Wang}}, \bibinfo {author}
  {\bibfnamefont {J.}~\bibnamefont {He}}, \bibinfo {author} {\bibfnamefont
  {X.}~\bibnamefont {Wang}}, \bibinfo {author} {\bibfnamefont {Y.}~\bibnamefont
  {Zhang}}, \ and\ \bibinfo {author} {\bibfnamefont {Z.}~\bibnamefont {Ye}},\
  }\href@noop {} {\bibfield  {journal} {\bibinfo  {journal} {Nat. Commun.}\
  }\textbf {\bibinfo {volume} {7}},\ \bibinfo {pages} {10896} (\bibinfo {year}
  {2016})}\BibitemShut {NoStop}%
\bibitem [{\citenamefont {Diroll}\ \emph {et~al.}(2017)\citenamefont {Diroll},
  \citenamefont {Nedelcu}, \citenamefont {Kovalenko},\ and\ \citenamefont
  {Schaller}}]{Diroll_AFM_27_2017}%
  \BibitemOpen
  \bibfield  {author} {\bibinfo {author} {\bibfnamefont {B.~T.}\ \bibnamefont
  {Diroll}}, \bibinfo {author} {\bibfnamefont {G.}~\bibnamefont {Nedelcu}},
  \bibinfo {author} {\bibfnamefont {M.~V.}\ \bibnamefont {Kovalenko}}, \ and\
  \bibinfo {author} {\bibfnamefont {R.~D.}\ \bibnamefont {Schaller}},\ }\href
  {\doibase 10.1002/adfm.201606750} {\bibfield  {journal} {\bibinfo  {journal}
  {Adv. Funct. Mater.}\ }\textbf {\bibinfo {volume} {27}},\ \bibinfo {pages}
  {1606750} (\bibinfo {year} {2017})}\BibitemShut {NoStop}%
\bibitem [{\citenamefont {Jadczak}\ \emph {et~al.}(2017)\citenamefont
  {Jadczak}, \citenamefont {Kutrowska-Girzycka}, \citenamefont
  {Kapu{\'{s}}ci{\'{n}}ski}, \citenamefont {Huang}, \citenamefont
  {W{\'{o}}js},\ and\ \citenamefont {Bryja}}]{Jadczak_N_28_2017}%
  \BibitemOpen
  \bibfield  {author} {\bibinfo {author} {\bibfnamefont {J.}~\bibnamefont
  {Jadczak}}, \bibinfo {author} {\bibfnamefont {J.}~\bibnamefont
  {Kutrowska-Girzycka}}, \bibinfo {author} {\bibfnamefont {P.}~\bibnamefont
  {Kapu{\'{s}}ci{\'{n}}ski}}, \bibinfo {author} {\bibfnamefont {Y.~S.}\
  \bibnamefont {Huang}}, \bibinfo {author} {\bibfnamefont {A.}~\bibnamefont
  {W{\'{o}}js}}, \ and\ \bibinfo {author} {\bibfnamefont {L.}~\bibnamefont
  {Bryja}},\ }\href {\doibase 10.1088/1361-6528/aa87d0} {\bibfield  {journal}
  {\bibinfo  {journal} {Nanotechnology}\ }\textbf {\bibinfo {volume} {28}},\
  \bibinfo {pages} {395702} (\bibinfo {year} {2017})}\BibitemShut {NoStop}%
\bibitem [{\citenamefont {Wilson}\ \emph {et~al.}(2018)\citenamefont {Wilson},
  \citenamefont {Hylton}, \citenamefont {Harada}, \citenamefont {Pearce},
  \citenamefont {Alonso-{\'{A}}lvarez}, \citenamefont {Mellor}, \citenamefont
  {Richards}, \citenamefont {David},\ and\ \citenamefont
  {Ekins-Daukes}}]{Wilson_SR_8_2018}%
  \BibitemOpen
  \bibfield  {author} {\bibinfo {author} {\bibfnamefont {T.}~\bibnamefont
  {Wilson}}, \bibinfo {author} {\bibfnamefont {N.~P.}\ \bibnamefont {Hylton}},
  \bibinfo {author} {\bibfnamefont {Y.}~\bibnamefont {Harada}}, \bibinfo
  {author} {\bibfnamefont {P.}~\bibnamefont {Pearce}}, \bibinfo {author}
  {\bibfnamefont {D.}~\bibnamefont {Alonso-{\'{A}}lvarez}}, \bibinfo {author}
  {\bibfnamefont {A.}~\bibnamefont {Mellor}}, \bibinfo {author} {\bibfnamefont
  {R.~D.}\ \bibnamefont {Richards}}, \bibinfo {author} {\bibfnamefont
  {J.~P.~R.}\ \bibnamefont {David}}, \ and\ \bibinfo {author} {\bibfnamefont
  {N.~J.}\ \bibnamefont {Ekins-Daukes}},\ }\href {\doibase
  10.1038/s41598-018-24696-2} {\bibfield  {journal} {\bibinfo  {journal} {Sci.
  Rep.}\ }\textbf {\bibinfo {volume} {8}} (\bibinfo {year} {2018}),\
  10.1038/s41598-018-24696-2}\BibitemShut {NoStop}%
\bibitem [{\citenamefont {Hidouri}\ \emph {et~al.}(2019)\citenamefont
  {Hidouri}, \citenamefont {Mal}, \citenamefont {Samajdar}, \citenamefont
  {Saidi},\ and\ \citenamefont {Das}}]{Hidouri_SM_129_2019}%
  \BibitemOpen
  \bibfield  {author} {\bibinfo {author} {\bibfnamefont {T.}~\bibnamefont
  {Hidouri}}, \bibinfo {author} {\bibfnamefont {I.}~\bibnamefont {Mal}},
  \bibinfo {author} {\bibfnamefont {D.~P.}\ \bibnamefont {Samajdar}}, \bibinfo
  {author} {\bibfnamefont {F.}~\bibnamefont {Saidi}}, \ and\ \bibinfo {author}
  {\bibfnamefont {T.~D.}\ \bibnamefont {Das}},\ }\href {\doibase
  10.1016/j.spmi.2019.04.003} {\bibfield  {journal} {\bibinfo  {journal}
  {Superlattices Microstruct.}\ }\textbf {\bibinfo {volume} {129}},\ \bibinfo
  {pages} {252} (\bibinfo {year} {2019})}\BibitemShut {NoStop}%
\bibitem [{\citenamefont {Reshchikov}\ \emph {et~al.}(2018)\citenamefont
  {Reshchikov}, \citenamefont {Ghimire},\ and\ \citenamefont
  {Demchenko}}]{Reshchikov_PRB_97_2018}%
  \BibitemOpen
  \bibfield  {author} {\bibinfo {author} {\bibfnamefont {M.~A.}\ \bibnamefont
  {Reshchikov}}, \bibinfo {author} {\bibfnamefont {P.}~\bibnamefont {Ghimire}},
  \ and\ \bibinfo {author} {\bibfnamefont {D.~O.}\ \bibnamefont {Demchenko}},\
  }\href {\doibase 10.1103/physrevb.97.205204} {\bibfield  {journal} {\bibinfo
  {journal} {Phys. Rev. B}\ }\textbf {\bibinfo {volume} {97}},\ \bibinfo
  {pages} {205204} (\bibinfo {year} {2018})}\BibitemShut {NoStop}%
\bibitem [{\citenamefont {Shockley}\ and\ \citenamefont
  {Read}(1952)}]{Shockley_PR_87_1952}%
  \BibitemOpen
  \bibfield  {author} {\bibinfo {author} {\bibfnamefont {W.}~\bibnamefont
  {Shockley}}\ and\ \bibinfo {author} {\bibfnamefont {W.~T.}\ \bibnamefont
  {Read}},\ }\href {\doibase 10.1103/PhysRev.87.835} {\bibfield  {journal}
  {\bibinfo  {journal} {Phys. Rev.}\ }\textbf {\bibinfo {volume} {87}},\
  \bibinfo {pages} {835} (\bibinfo {year} {1952})}\BibitemShut {NoStop}%
\bibitem [{\citenamefont {Hall}(1952)}]{Hall_PR_87_1952}%
  \BibitemOpen
  \bibfield  {author} {\bibinfo {author} {\bibfnamefont {R.~N.}\ \bibnamefont
  {Hall}},\ }\href {\doibase 10.1103/PhysRev.87.387} {\bibfield  {journal}
  {\bibinfo  {journal} {Phys. Rev.}\ }\textbf {\bibinfo {volume} {87}},\
  \bibinfo {pages} {387} (\bibinfo {year} {1952})}\BibitemShut {NoStop}%
\bibitem [{\citenamefont {Li}\ \emph {et~al.}(2017)\citenamefont {Li},
  \citenamefont {Piccardo}, \citenamefont {Lu}, \citenamefont {Mayboroda},
  \citenamefont {Martinelli}, \citenamefont {Peretti}, \citenamefont {Speck},
  \citenamefont {Weisbuch}, \citenamefont {Filoche},\ and\ \citenamefont
  {Wu}}]{Li_PRB_95_2017}%
  \BibitemOpen
  \bibfield  {author} {\bibinfo {author} {\bibfnamefont {C.-K.}\ \bibnamefont
  {Li}}, \bibinfo {author} {\bibfnamefont {M.}~\bibnamefont {Piccardo}},
  \bibinfo {author} {\bibfnamefont {L.-S.}\ \bibnamefont {Lu}}, \bibinfo
  {author} {\bibfnamefont {S.}~\bibnamefont {Mayboroda}}, \bibinfo {author}
  {\bibfnamefont {L.}~\bibnamefont {Martinelli}}, \bibinfo {author}
  {\bibfnamefont {J.}~\bibnamefont {Peretti}}, \bibinfo {author} {\bibfnamefont
  {J.~S.}\ \bibnamefont {Speck}}, \bibinfo {author} {\bibfnamefont
  {C.}~\bibnamefont {Weisbuch}}, \bibinfo {author} {\bibfnamefont
  {M.}~\bibnamefont {Filoche}}, \ and\ \bibinfo {author} {\bibfnamefont
  {Y.-R.}\ \bibnamefont {Wu}},\ }\href {\doibase 10.1103/PhysRevB.95.144206}
  {\bibfield  {journal} {\bibinfo  {journal} {Phys. Rev. B}\ }\textbf {\bibinfo
  {volume} {95}},\ \bibinfo {pages} {144206} (\bibinfo {year}
  {2017})}\BibitemShut {NoStop}%
\bibitem [{\citenamefont {Gee}\ and\ \citenamefont
  {Kastner}(1979)}]{Gee_PRL_42_1979}%
  \BibitemOpen
  \bibfield  {author} {\bibinfo {author} {\bibfnamefont {C.~M.}\ \bibnamefont
  {Gee}}\ and\ \bibinfo {author} {\bibfnamefont {M.}~\bibnamefont {Kastner}},\
  }\href@noop {} {\bibfield  {journal} {\bibinfo  {journal} {Phys. Rev. Lett.}\
  }\textbf {\bibinfo {volume} {42}},\ \bibinfo {pages} {1765} (\bibinfo {year}
  {1979})}\BibitemShut {NoStop}%
\bibitem [{\citenamefont {Rubel}\ \emph {et~al.}(2006)\citenamefont {Rubel},
  \citenamefont {Baranovskii}, \citenamefont {Hantke}, \citenamefont {Kunert},
  \citenamefont {Ruhle}, \citenamefont {Thomas}, \citenamefont {Volz},\ and\
  \citenamefont {Stolz}}]{Rubel_PRB_73_2006}%
  \BibitemOpen
  \bibfield  {author} {\bibinfo {author} {\bibfnamefont {O.}~\bibnamefont
  {Rubel}}, \bibinfo {author} {\bibfnamefont {S.~D.}\ \bibnamefont
  {Baranovskii}}, \bibinfo {author} {\bibfnamefont {K.}~\bibnamefont {Hantke}},
  \bibinfo {author} {\bibfnamefont {B.}~\bibnamefont {Kunert}}, \bibinfo
  {author} {\bibfnamefont {W.~W.}\ \bibnamefont {Ruhle}}, \bibinfo {author}
  {\bibfnamefont {P.}~\bibnamefont {Thomas}}, \bibinfo {author} {\bibfnamefont
  {K.}~\bibnamefont {Volz}}, \ and\ \bibinfo {author} {\bibfnamefont
  {W.}~\bibnamefont {Stolz}},\ }\href {\doibase 10.1103/PhysRevB.73.233201}
  {\bibfield  {journal} {\bibinfo  {journal} {Phys. Rev. B}\ }\textbf {\bibinfo
  {volume} {73}},\ \bibinfo {eid} {233201} (\bibinfo {year}
  {2006})}\BibitemShut {NoStop}%
\bibitem [{\citenamefont {Shakfa}\ \emph {et~al.}(2015)\citenamefont {Shakfa},
  \citenamefont {Wiemer}, \citenamefont {Ludewig}, \citenamefont {Jandieri},
  \citenamefont {Volz}, \citenamefont {Stolz}, \citenamefont {Baranovskii},\
  and\ \citenamefont {Koch}}]{Shakfa_JAP_117_2015}%
  \BibitemOpen
  \bibfield  {author} {\bibinfo {author} {\bibfnamefont {M.~K.}\ \bibnamefont
  {Shakfa}}, \bibinfo {author} {\bibfnamefont {M.}~\bibnamefont {Wiemer}},
  \bibinfo {author} {\bibfnamefont {P.}~\bibnamefont {Ludewig}}, \bibinfo
  {author} {\bibfnamefont {K.}~\bibnamefont {Jandieri}}, \bibinfo {author}
  {\bibfnamefont {K.}~\bibnamefont {Volz}}, \bibinfo {author} {\bibfnamefont
  {W.}~\bibnamefont {Stolz}}, \bibinfo {author} {\bibfnamefont {S.~D.}\
  \bibnamefont {Baranovskii}}, \ and\ \bibinfo {author} {\bibfnamefont
  {M.}~\bibnamefont {Koch}},\ }\href@noop {} {\bibfield  {journal} {\bibinfo
  {journal} {J. Appl. Phys.}\ }\textbf {\bibinfo {volume} {117}},\ \bibinfo
  {pages} {025709} (\bibinfo {year} {2015})}\BibitemShut {NoStop}%
\bibitem [{\citenamefont {Jandieri}\ \emph {et~al.}(2013)\citenamefont
  {Jandieri}, \citenamefont {Kunert}, \citenamefont {Liebich}, \citenamefont
  {Zimprich}, \citenamefont {Volz}, \citenamefont {Stolz}, \citenamefont
  {Gebhard}, \citenamefont {Baranovskii}, \citenamefont {Koukourakis},
  \citenamefont {Gerhardt},\ and\ \citenamefont
  {Hofmann}}]{Jandieri_PRB_87_2013}%
  \BibitemOpen
  \bibfield  {author} {\bibinfo {author} {\bibfnamefont {K.}~\bibnamefont
  {Jandieri}}, \bibinfo {author} {\bibfnamefont {B.}~\bibnamefont {Kunert}},
  \bibinfo {author} {\bibfnamefont {S.}~\bibnamefont {Liebich}}, \bibinfo
  {author} {\bibfnamefont {M.}~\bibnamefont {Zimprich}}, \bibinfo {author}
  {\bibfnamefont {K.}~\bibnamefont {Volz}}, \bibinfo {author} {\bibfnamefont
  {W.}~\bibnamefont {Stolz}}, \bibinfo {author} {\bibfnamefont
  {F.}~\bibnamefont {Gebhard}}, \bibinfo {author} {\bibfnamefont {S.~D.}\
  \bibnamefont {Baranovskii}}, \bibinfo {author} {\bibfnamefont
  {N.}~\bibnamefont {Koukourakis}}, \bibinfo {author} {\bibfnamefont {N.~C.}\
  \bibnamefont {Gerhardt}}, \ and\ \bibinfo {author} {\bibfnamefont {M.~R.}\
  \bibnamefont {Hofmann}},\ }\href {\doibase 10.1103/PhysRevB.87.035303}
  {\bibfield  {journal} {\bibinfo  {journal} {Phys. Rev. B}\ }\textbf {\bibinfo
  {volume} {87}},\ \bibinfo {pages} {035303} (\bibinfo {year}
  {2013})}\BibitemShut {NoStop}%
\bibitem [{\citenamefont {Lambkin}\ \emph {et~al.}(1994)\citenamefont
  {Lambkin}, \citenamefont {Considine}, \citenamefont {Walsh}, \citenamefont
  {O'connor}, \citenamefont {McDonagh},\ and\ \citenamefont
  {Glynn}}]{Lambkin_APL_65_1994}%
  \BibitemOpen
  \bibfield  {author} {\bibinfo {author} {\bibfnamefont {J.~D.}\ \bibnamefont
  {Lambkin}}, \bibinfo {author} {\bibfnamefont {L.}~\bibnamefont {Considine}},
  \bibinfo {author} {\bibfnamefont {S.}~\bibnamefont {Walsh}}, \bibinfo
  {author} {\bibfnamefont {G.~M.}\ \bibnamefont {O'connor}}, \bibinfo {author}
  {\bibfnamefont {C.~J.}\ \bibnamefont {McDonagh}}, \ and\ \bibinfo {author}
  {\bibfnamefont {T.~J.}\ \bibnamefont {Glynn}},\ }\href@noop {} {\bibfield
  {journal} {\bibinfo  {journal} {Appl. Phys. Lett.}\ }\textbf {\bibinfo
  {volume} {65}},\ \bibinfo {pages} {73} (\bibinfo {year} {1994})}\BibitemShut
  {NoStop}%
\bibitem [{\citenamefont {Sun}\ \emph {et~al.}(2006)\citenamefont {Sun},
  \citenamefont {Calvez}, \citenamefont {Dawson}, \citenamefont {Gupta},
  \citenamefont {Aers},\ and\ \citenamefont {Sproule}}]{Sun_APL_89_2006}%
  \BibitemOpen
  \bibfield  {author} {\bibinfo {author} {\bibfnamefont {H.~D.}\ \bibnamefont
  {Sun}}, \bibinfo {author} {\bibfnamefont {S.}~\bibnamefont {Calvez}},
  \bibinfo {author} {\bibfnamefont {M.~D.}\ \bibnamefont {Dawson}}, \bibinfo
  {author} {\bibfnamefont {J.~A.}\ \bibnamefont {Gupta}}, \bibinfo {author}
  {\bibfnamefont {G.~C.}\ \bibnamefont {Aers}}, \ and\ \bibinfo {author}
  {\bibfnamefont {G.~I.}\ \bibnamefont {Sproule}},\ }\href@noop {} {\bibfield
  {journal} {\bibinfo  {journal} {Appl. Phys. Lett.}\ }\textbf {\bibinfo
  {volume} {89}},\ \bibinfo {pages} {101909} (\bibinfo {year}
  {2006})}\BibitemShut {NoStop}%
\bibitem [{\citenamefont {Stoneham}(1981)}]{Stoneham_RPP_44_1981}%
  \BibitemOpen
  \bibfield  {author} {\bibinfo {author} {\bibfnamefont {A.~M.}\ \bibnamefont
  {Stoneham}},\ }\href {\doibase 10.1088/0034-4885/44/12/001} {\bibfield
  {journal} {\bibinfo  {journal} {Rep. Prog. Phys.}\ }\textbf {\bibinfo
  {volume} {44}},\ \bibinfo {pages} {1251} (\bibinfo {year}
  {1981})}\BibitemShut {NoStop}%
\bibitem [{\citenamefont {Daly}\ \emph {et~al.}(1995)\citenamefont {Daly},
  \citenamefont {Glynn}, \citenamefont {Lambkin}, \citenamefont {Considine},\
  and\ \citenamefont {Walsh}}]{Daly_PRB_52_1995}%
  \BibitemOpen
  \bibfield  {author} {\bibinfo {author} {\bibfnamefont {E.~M.}\ \bibnamefont
  {Daly}}, \bibinfo {author} {\bibfnamefont {T.~J.}\ \bibnamefont {Glynn}},
  \bibinfo {author} {\bibfnamefont {J.~D.}\ \bibnamefont {Lambkin}}, \bibinfo
  {author} {\bibfnamefont {L.}~\bibnamefont {Considine}}, \ and\ \bibinfo
  {author} {\bibfnamefont {S.}~\bibnamefont {Walsh}},\ }\href {\doibase
  10.1103/PhysRevB.52.4696} {\bibfield  {journal} {\bibinfo  {journal} {Phys.
  Rev. B}\ }\textbf {\bibinfo {volume} {52}},\ \bibinfo {pages} {4696}
  (\bibinfo {year} {1995})}\BibitemShut {NoStop}%
\bibitem [{\citenamefont {Buyanova}\ \emph {et~al.}(2003)\citenamefont
  {Buyanova}, \citenamefont {Izadifard}, \citenamefont {Chen}, \citenamefont
  {Polimeni}, \citenamefont {Capizzi}, \citenamefont {Xin},\ and\ \citenamefont
  {Tu}}]{Buyanova_APL_82_2003}%
  \BibitemOpen
  \bibfield  {author} {\bibinfo {author} {\bibfnamefont {I.~A.}\ \bibnamefont
  {Buyanova}}, \bibinfo {author} {\bibfnamefont {M.}~\bibnamefont {Izadifard}},
  \bibinfo {author} {\bibfnamefont {W.~M.}\ \bibnamefont {Chen}}, \bibinfo
  {author} {\bibfnamefont {A.}~\bibnamefont {Polimeni}}, \bibinfo {author}
  {\bibfnamefont {M.}~\bibnamefont {Capizzi}}, \bibinfo {author} {\bibfnamefont
  {H.~P.}\ \bibnamefont {Xin}}, \ and\ \bibinfo {author} {\bibfnamefont
  {C.~W.}\ \bibnamefont {Tu}},\ }\href@noop {} {\bibfield  {journal} {\bibinfo
  {journal} {Appl. Phys. Lett.}\ }\textbf {\bibinfo {volume} {82}},\ \bibinfo
  {pages} {3662} (\bibinfo {year} {2003})}\BibitemShut {NoStop}%
\bibitem [{\citenamefont {Wang}\ \emph {et~al.}(2001)\citenamefont {Wang},
  \citenamefont {Yoon}, \citenamefont {He},\ and\ \citenamefont
  {Shen}}]{Wang_JAP_90_2001}%
  \BibitemOpen
  \bibfield  {author} {\bibinfo {author} {\bibfnamefont {S.~Z.}\ \bibnamefont
  {Wang}}, \bibinfo {author} {\bibfnamefont {S.~F.}\ \bibnamefont {Yoon}},
  \bibinfo {author} {\bibfnamefont {L.}~\bibnamefont {He}}, \ and\ \bibinfo
  {author} {\bibfnamefont {X.~C.}\ \bibnamefont {Shen}},\ }\href@noop {}
  {\bibfield  {journal} {\bibinfo  {journal} {J. Appl. Phys.}\ }\textbf
  {\bibinfo {volume} {90}},\ \bibinfo {pages} {2314} (\bibinfo {year}
  {2001})}\BibitemShut {NoStop}%
\bibitem [{\citenamefont {Lourenco}\ \emph {et~al.}(2003)\citenamefont
  {Lourenco}, \citenamefont {Dias}, \citenamefont {Pocas}, \citenamefont
  {Duarte}, \citenamefont {De~Oliveira},\ and\ \citenamefont
  {Harmand}}]{Lourenco_JAP_93_2003}%
  \BibitemOpen
  \bibfield  {author} {\bibinfo {author} {\bibfnamefont {S.~A.}\ \bibnamefont
  {Lourenco}}, \bibinfo {author} {\bibfnamefont {I.~F.~L.}\ \bibnamefont
  {Dias}}, \bibinfo {author} {\bibfnamefont {L.~C.}\ \bibnamefont {Pocas}},
  \bibinfo {author} {\bibfnamefont {J.~L.}\ \bibnamefont {Duarte}}, \bibinfo
  {author} {\bibfnamefont {J.~B.~B.}\ \bibnamefont {De~Oliveira}}, \ and\
  \bibinfo {author} {\bibfnamefont {J.~C.}\ \bibnamefont {Harmand}},\
  }\href@noop {} {\bibfield  {journal} {\bibinfo  {journal} {J. Appl. Phys.}\
  }\textbf {\bibinfo {volume} {93}},\ \bibinfo {pages} {4475} (\bibinfo {year}
  {2003})}\BibitemShut {NoStop}%
\bibitem [{\citenamefont {Rubel}\ \emph {et~al.}(2005)\citenamefont {Rubel},
  \citenamefont {Galluppi}, \citenamefont {Baranovskii}, \citenamefont {Volz},
  \citenamefont {Geelhaar}, \citenamefont {Riechert}, \citenamefont {Thomas},\
  and\ \citenamefont {Stolz}}]{Rubel_JAP_98_2005}%
  \BibitemOpen
  \bibfield  {author} {\bibinfo {author} {\bibfnamefont {O.}~\bibnamefont
  {Rubel}}, \bibinfo {author} {\bibfnamefont {M.}~\bibnamefont {Galluppi}},
  \bibinfo {author} {\bibfnamefont {S.~D.}\ \bibnamefont {Baranovskii}},
  \bibinfo {author} {\bibfnamefont {K.}~\bibnamefont {Volz}}, \bibinfo {author}
  {\bibfnamefont {L.}~\bibnamefont {Geelhaar}}, \bibinfo {author}
  {\bibfnamefont {H.}~\bibnamefont {Riechert}}, \bibinfo {author}
  {\bibfnamefont {P.}~\bibnamefont {Thomas}}, \ and\ \bibinfo {author}
  {\bibfnamefont {W.}~\bibnamefont {Stolz}},\ }\href@noop {} {\bibfield
  {journal} {\bibinfo  {journal} {J. Appl. Phys.}\ }\textbf {\bibinfo {volume}
  {98}},\ \bibinfo {pages} {063518} (\bibinfo {year} {2005})}\BibitemShut
  {NoStop}%
\bibitem [{\citenamefont {Borkovska}\ \emph {et~al.}(2006)\citenamefont
  {Borkovska}, \citenamefont {Yefanov}, \citenamefont {Gudymenko},
  \citenamefont {Johnson}, \citenamefont {Kladko}, \citenamefont {Korsunska},
  \citenamefont {Kryshtab}, \citenamefont {Sadofyev},\ and\ \citenamefont
  {Zhang}}]{Borkovska_TSF_515_2006}%
  \BibitemOpen
  \bibfield  {author} {\bibinfo {author} {\bibfnamefont {L.}~\bibnamefont
  {Borkovska}}, \bibinfo {author} {\bibfnamefont {O.}~\bibnamefont {Yefanov}},
  \bibinfo {author} {\bibfnamefont {O.}~\bibnamefont {Gudymenko}}, \bibinfo
  {author} {\bibfnamefont {S.}~\bibnamefont {Johnson}}, \bibinfo {author}
  {\bibfnamefont {V.}~\bibnamefont {Kladko}}, \bibinfo {author} {\bibfnamefont
  {N.}~\bibnamefont {Korsunska}}, \bibinfo {author} {\bibfnamefont
  {T.}~\bibnamefont {Kryshtab}}, \bibinfo {author} {\bibfnamefont
  {Y.}~\bibnamefont {Sadofyev}}, \ and\ \bibinfo {author} {\bibfnamefont
  {Y.-H.}\ \bibnamefont {Zhang}},\ }\href {\doibase 10.1016/j.tsf.2005.12.194}
  {\bibfield  {journal} {\bibinfo  {journal} {Thin Solid Films}\ }\textbf
  {\bibinfo {volume} {515}},\ \bibinfo {pages} {786} (\bibinfo {year}
  {2006})}\BibitemShut {NoStop}%
\bibitem [{\citenamefont {Mazur}\ \emph {et~al.}(2013)\citenamefont {Mazur},
  \citenamefont {Dorogan}, \citenamefont {Schmidbauer}, \citenamefont
  {Tarasov}, \citenamefont {Johnson}, \citenamefont {Lu}, \citenamefont {Ware},
  \citenamefont {Yu}, \citenamefont {Tiedje},\ and\ \citenamefont
  {Salamo}}]{Mazur_JAP_113_2013}%
  \BibitemOpen
  \bibfield  {author} {\bibinfo {author} {\bibfnamefont {Y.~I.}\ \bibnamefont
  {Mazur}}, \bibinfo {author} {\bibfnamefont {V.~G.}\ \bibnamefont {Dorogan}},
  \bibinfo {author} {\bibfnamefont {M.}~\bibnamefont {Schmidbauer}}, \bibinfo
  {author} {\bibfnamefont {G.~G.}\ \bibnamefont {Tarasov}}, \bibinfo {author}
  {\bibfnamefont {S.~R.}\ \bibnamefont {Johnson}}, \bibinfo {author}
  {\bibfnamefont {X.}~\bibnamefont {Lu}}, \bibinfo {author} {\bibfnamefont
  {M.~E.}\ \bibnamefont {Ware}}, \bibinfo {author} {\bibfnamefont {S.-Q.}\
  \bibnamefont {Yu}}, \bibinfo {author} {\bibfnamefont {T.}~\bibnamefont
  {Tiedje}}, \ and\ \bibinfo {author} {\bibfnamefont {G.~J.}\ \bibnamefont
  {Salamo}},\ }\href {\doibase 10.1063/1.4801429} {\bibfield  {journal}
  {\bibinfo  {journal} {J. Appl. Phys.}\ }\textbf {\bibinfo {volume} {113}},\
  \bibinfo {pages} {144308} (\bibinfo {year} {2013})}\BibitemShut {NoStop}%
\bibitem [{\citenamefont {Linhart}\ \emph {et~al.}(2017)\citenamefont
  {Linhart}, \citenamefont {Gladysiewicz}, \citenamefont {Kopaczek},
  \citenamefont {Rajpalke}, \citenamefont {Ashwin}, \citenamefont {Veal},\ and\
  \citenamefont {Kudrawiec}}]{Linhart_JPDAP_50_2017}%
  \BibitemOpen
  \bibfield  {author} {\bibinfo {author} {\bibfnamefont {W.~M.}\ \bibnamefont
  {Linhart}}, \bibinfo {author} {\bibfnamefont {M.}~\bibnamefont
  {Gladysiewicz}}, \bibinfo {author} {\bibfnamefont {J.}~\bibnamefont
  {Kopaczek}}, \bibinfo {author} {\bibfnamefont {M.~K.}\ \bibnamefont
  {Rajpalke}}, \bibinfo {author} {\bibfnamefont {M.~J.}\ \bibnamefont
  {Ashwin}}, \bibinfo {author} {\bibfnamefont {T.~D.}\ \bibnamefont {Veal}}, \
  and\ \bibinfo {author} {\bibfnamefont {R.}~\bibnamefont {Kudrawiec}},\ }\href
  {\doibase 10.1088/1361-6463/aa7e64} {\bibfield  {journal} {\bibinfo
  {journal} {J. Phys. D: Appl. Phys.}\ }\textbf {\bibinfo {volume} {50}},\
  \bibinfo {pages} {375102} (\bibinfo {year} {2017})}\BibitemShut {NoStop}%
\bibitem [{\citenamefont {Schmidt}\ \emph {et~al.}(1992)\citenamefont
  {Schmidt}, \citenamefont {Lischka},\ and\ \citenamefont
  {Zulehner}}]{Schmidt_PRB_45_1992}%
  \BibitemOpen
  \bibfield  {author} {\bibinfo {author} {\bibfnamefont {T.}~\bibnamefont
  {Schmidt}}, \bibinfo {author} {\bibfnamefont {K.}~\bibnamefont {Lischka}}, \
  and\ \bibinfo {author} {\bibfnamefont {W.}~\bibnamefont {Zulehner}},\ }\href
  {\doibase 10.1103/physrevb.45.8989} {\bibfield  {journal} {\bibinfo
  {journal} {Phys. Rev. B}\ }\textbf {\bibinfo {volume} {45}},\ \bibinfo
  {pages} {8989} (\bibinfo {year} {1992})}\BibitemShut {NoStop}%
\end{thebibliography}

%

%
%
\newpage
\onecolumngrid

\begin{table}[h]
  \caption{List of material parameters that enter the model. Temperature-dependent quantities are marked with $(T)$. The degree symbol ($^\circ$) indicates a value of material parameters taken at standard conditions ($T^\circ = 300$~K).}\label{Table:1}
  \begin{ruledtabular}
    \begin{tabular}{l l l l}
      Symbol & Description & Value & Units \\
      \hline
      $E_l$ & Localization energy relative to $E_c$ (Fig.~\ref{Fig:Proc-schem}) & $0.02$ & eV \\
      $N_c(T)$ & Effective density of extended states in the conduction band & $4.5 \times 10^{17} (T/T^\circ)^{3/2}$ & cm$^{-3}$ \\
      $N_v(T)$ & Effective density of extended states in the valence band & $20 \, N_c(T)$ & cm$^{-3}$ \\
      $N_l$ & Density of localized states & $(2/5) \, N_c^\circ$ & cm$^{-3}$ \\
      $N_t$ & Density of deep traps & $N_l/10$ & cm$^{-3}$ \\
      $\sigma_n$ & Electron capture cross section for localized states and deep traps\footnote{The same cross section is assumed for localized states and deep traps to make the algebra more simple.} & $5 \times 10^{-13}$ & cm$^{2}$ \\
      $v_\text{th}(T)$ & Electron thermal velocity & $4.4 \times 10^5 (T/T^\circ)^{1/2}$ & cm~s$^{-1}$ \\
      $C_n(T)$ & Electron capture coefficient for localized states and deep traps & $\sigma_n v_\text{th}(T)$ & cm$^{3}$~s$^{-1}$ \\
      $C_p(T)$ & Hole capture coefficient for deep traps & $C_n(T)/10$ & cm$^{3}$~s$^{-1}$ \\
      $B_{cv}$ & Bimolecular recombination coefficient for extended states\footnote{The bimolecular recombination coefficient is assumed temperature independent since a functional form of this dependence is unknown. However, it generally shows a strong temperature dependence with a tendency to decrease with decreasing temperature.} & $7 \times 10^{-10}$ & cm$^{3}$~s$^{-1}$ \\
      $B_{lv}$ & Bimolecular recombination coefficient for localized states & $B_{cv}/10$ & cm$^{3}$~s$^{-1}$ \\
    \end{tabular}
  \end{ruledtabular}
\end{table}

%
%
\clearpage

\end{document}